\documentclass[runningheads]{cl2emult}

\usepackage{makeidx}  % allows index generation
\usepackage{graphicx} % standard LaTeX graphics tool
\begin{document}

\title*{The Hubble Constant from the Fornax~Cluster~Distance
  \thanks{to be published in: Science in the VLT Era and Beyond. ESO
    VLT Opening Symposium, 1-4 March 1999, Antofagasta, Chile.
    Springer-Verlag, Berlin, in press}}
\index{globular~cluster}\index{supernova}\index{Hubble~constant}

\author{Tom~Richtler\inst{1} \and Georg~Drenkhahn\inst{2,1},
  Mat\'\i as~G\'omez\inst{1,3} \and Wilhelm~Seggewiss\inst{1}}

\authorrunning{Tom Richtler et~al.}

\institute{Sternwarte der Universit\"at Bonn, Auf dem H\"ugel 71,
  D-53121 Bonn, Germany \and Max-Planck-Institut f\"ur Astrophysik,
  D-85740 Garching, Germany \and Departamento de Astronom\'\i a y
  Astrof\'\i sica, PUC, Santiago de Chile}

\maketitle 

\begin{abstract}
  Type Ia supernovae are the best cosmological standard candles
  available.  The intrinsic scatter of their decline-rate- and
  colour-corrected peak brightnesses in the Hubble diagram is within
  observational error limits, corresponding to an uncertainty of only
  3\,km\,s$^{-1}$\,Mpc$^{-1}$ of the Hubble constant. Any additional
  uncertainty, resulting from peak-brightness calibration, must be
  kept small by measuring distances to nearby host galaxies most
  precisely.
  
  A number of different distance determinations of the Fornax cluster
  of galaxies agree well on a distance modulus of $31.35\pm0.04$\,mag
  ($18.6\pm0.3$\,Mpc). This leads to accurate absolute magnitudes of
  the well-observed Fornax type Ia SNe SN\,1980N, SN\,1981D, and
  SN\,1992A and finally to a Hubble constant of\\
  $H_0=72\pm6$\,km\,s$^{-1}$\,Mpc$^{-1}$.
\end{abstract}

\section{The Hubble diagram of type Ia SNe}

Supernovae (SNe) of type Ia\index{supernova!type~Ia} are known to be
excellent standard candles.  In order to construct a reliable Hubble
diagram\index{Hubble~diagram} it needs more than just to plot the data
from the standard SNe catalogue \cite{barbon:89} but requires highly
uniform spectroscopic and, especially, photometric observations as are
now available from the Cal\'an-Tololo survey \cite{hamuy:96} or the
CfA campaign \cite{riess:99}.

\begin{figure}
  \centering
  \includegraphics[height=7.1cm]{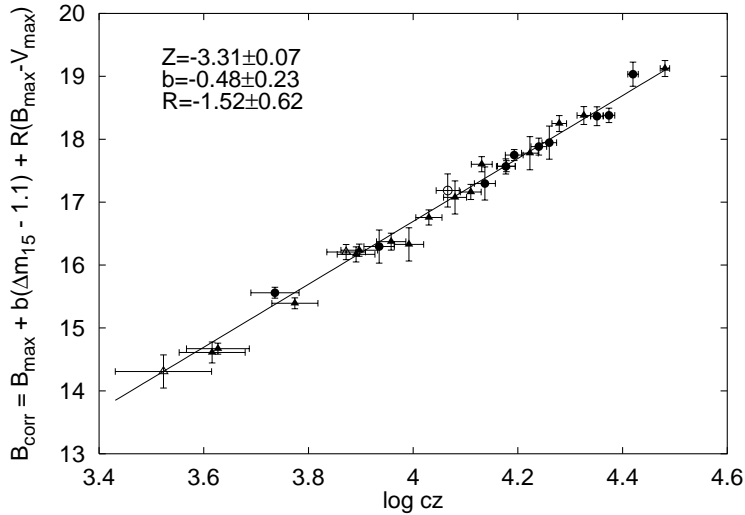}
  \caption{The Hubble diagram of corrected SN Ia magnitudes in $B$
    based on the C\'alan-Tololo sample \cite{hamuy:96}. $b$ and $R$
    are the correction coefficients for decline rate and intrinsic
    colour, resp., and $Z$ is the intersection of the straight line
    with the $y$-axis. No systematic differences are seen between
    SNe Ia's in early-type galaxies (circles), late-type galaxies
    (triangles), and ``red'' SNe (open symbols).}
  \label{Hamuy}
\end{figure}

Fig.~\ref{Hamuy} shows the Hubble diagram for 29 classified Ia SNe
from the Cal\'an-Tololo survey \cite{hamuy:96}. The intersection $Z$
of the straight line (slope 5!) with the $y$-axis is related to the
Hubble constant $H_0$ and the absolute magnitude $M$ of the standard
candles by the well-known equation:
\begin{equation}
  \label{eq:h0}
  \log H_0 = 0.2\cdot(M-Z)+5
\end{equation}
For the knowledge of an accurate Hubble constant, one would of course
aim at determining the distances to as many SNe as possible.

The quality of the SNe Ia Hubble diagram can be improved further if
one corrects the brightness at maximum for the decline rate
\index{supernova!type~Ia!decline~rate} $\Delta m_{15}$ of the $B$
lightcurve (e.g. \cite{hamuy:96}) and the colour
\index{supernova!type~Ia!colour} $(B_\mathrm{max}-V_\mathrm{max})$
\cite{tripp:98}:
\begin{equation}
  \label{eq:corr}
  B_\mathrm{corr} = B_\mathrm{max} + b(\Delta m_{15} - 1.1) 
  + R(B_\mathrm{max}-V_\mathrm{max})
\end{equation}
\index{supernova!type~Ia!corrected~magnitude} A new analysis of the
Cal\'an-Tololo SNe sample has been made recently by Drenkhahn \&
Richtler \cite{drenkhahn:99,richtler:99a}.  The derived correction
terms are quoted in the upper part of Fig.~\ref{Hamuy}.  As a result
of the correction process the scatter in the Hubble diagram is reduced
to 0.14\,mag (even smaller in $V$ and $I$), which is already
consistent with the observational errors.  Therefore, the intrinsic
scatter must be distinctly smaller than 0.1\,mag. This fixes the
zero-point $Z$ with high precision and the accuracy of $H_0$ in
(\ref{eq:h0}) depends only on the absolute magnitude
$M_\mathrm{corr}$.  Determining $M_\mathrm{corr}$ within $\pm0.1$\,mag
corresponds to measuring $H_0$ with an uncertainty of only
$\pm3$\,km\,s$^{-1}$\,Mpc$^{-1}$.

\section{The distance of the Fornax cluster of galaxies}

\index{Fornax~cluster} With respect to the accurate calibration of the
absolute magnitudes of Ia SNe, the Fornax cluster of galaxies is one
of the most suitable sites. It hosts the type Ia SNe SN\,1992A (in NGC\,1380),
SN\,1980N, and SN\,1981D (both in NGC\,1316).  In contrast to the
Virgo cluster\index{Virgo~cluster}, which is the permanent but
precarious target of distance measurements, the Fornax cluster has a
compact structure without distinct substructures.  Promising attempts
to tighten the distance to Fornax have been made
\cite{mcmillan:93,kohle:96,bureau:96}, but have been criticised when
hunting for a small value of the Hubble constant \cite{tammann:99}.
The last year has brought convincing new results for the Fornax
distance (for the following discussion refer to
Table~\ref{tab:fornax}).

\index{Fornax~cluster!distance}
\noindent\textit{Planetary nebula luminosity function (PNLF).}
\index{Planetary nebula luminosity function
  (PNLF)}\index{Fornax~cluster!distance!PNLF} Distances to Fornax
galaxies using the PNLF have been deduced by Jacoby and co-workers
\cite{mcmillan:93,jacoby:97}. Their zero-point is based on M31 with a
distance of 710\,kpc (or distance modulus $m - M = 24.26$\,mag) and a
reddening of $E_{B-V}=0.11$\,mag. Changing to the more recent distance
modulus of $m-M=24.44$\,mag \cite{federspiel:98} but keeping the
reddening one arrives at a Fornax modulus of $31.33\pm0.08$\,mag.

\sloppypar
\noindent\textit{Globular cluster luminosity function (GCLF).}
\index{GCLF}\index{Fornax~cluster!distance!GCLF} The turn-over
magnitude (TOM) of the GCLF of four Fornax galaxies has been derived
by Kohle et~al.  \cite{kohle:96} to be
$V_\mathrm{TO}=23.69\pm0.06$\,mag (error weighted and skipping
NGC\,1379). Applying a more recent calibration of the TOM of the
Galactic globular cluster system:
$M_{V,\mathrm{TO}}=-7.61\pm0.08$\,mag \cite{drenkhahn:99,tammann:99},
\index{GCLF!Galactic turn-over magnitude} one gets a Fornax distance
modulus of $31.30\pm0.13$\,mag.  The GCLF of
NGC\,1380\index{NGC\,1380} has also been analysed using ESO NTT data
\cite{della:98}, which result in a modulus of $31.35\pm0.16$\,mag.
Presently, we are investigating the GCLF of the Fornax galaxy
NGC\,1316\index{NGC\,1316} \cite{richtler:99}. The preliminary result
of $31.32\pm0.15$\,mag agrees well with the other values.

\noindent\textit{Cepheid distances.}
\index{Fornax~cluster!distance!Cepheids} Now there is also a Cepheid
distance of one Fornax galaxy, NGC\,1365, available (HST Key Project
\cite{madore:98}). The value of $31.35\pm0.07$\,mag agrees very well
with the results of the previously mentioned methods.

\noindent\textit{Surface brightness fluctuations (SBF).}
\index{Surface brightness fluctuations (SBF)}
\index{Fornax~cluster!distance!SBF} Jensen et~al. \cite{jensen:98}
report new infrared SBF measurements for five Fornax galaxies. They
find a distance modulus of $31.37\pm0.12$\,mag.

\noindent\textit{Tully-Fisher relation (T-F).}
\index{Tully-Fisher relation (T-F)}\index{Fornax~cluster!distance!T-F}
Finally, a Tully-Fisher distance from 21\,cm data for 18 Fornax
galaxies can be derived: when combining published 21\,cm line widths
$\Delta V_{20}$ \cite{bureau:96} with dereddened integrated $B$
magnitudes $B_\mathrm{tot}^0$ \cite{rc3:91} one finds the relation:
$B_\mathrm{tot}^0 = -6.990 (\pm 0.426) \log \Delta V_{20} + 29.312
(\pm 0.365)$ (N.B.: The galaxy ESO~357-G25 which has an extremely
large error in the line width has been skipped from the sample
\cite[their Table~1]{bureau:96}.)  We then use the absolute magnitudes
$M_B$ of all non-Fornax calibrating galaxies from the work of
Federspiel et~al. \cite[their Table~2]{federspiel:98} and obtain
$31.43\pm0.12$\,mag.

\index{Fornax~cluster!distance} Table~\ref{tab:fornax} suggests that
the distance of the Fornax cluster is now very well determined. The
mean value of the distance modulus is $31.35$\,mag with a scatter of
only $\pm0.04$\,mag (linear distance $18.6\pm0.3$\,Mpc).
 
\begin{table}
  \caption{Distance modulus determinations of the Fornax cluster}
  \setlength\tabcolsep{11pt}
  \label{tab:fornax}
  \begin{tabular}{lccll}
    \hline\noalign{\smallskip}
    Galaxy & $m - M$ & $\sigma$ & Method$^\mathrm{a}$ & Reference \\
    NGC & mag & mag & & \\
    \noalign{\smallskip}
    \hline
    \noalign{\smallskip}
    1316, 1399, 1404 & 31.33 & 0.08 
      & PNLF & \cite{mcmillan:93}, \cite{jacoby:97} \\
    1374, 1399, 1427 & 31.30 & 0.11
      & GCLF & \cite{kohle:96} \\
    1380 & 31.35 & 0.16 & GCLF & \cite{della:98} \\
    1316 & 31.32 & 0.15 & GCLF & \cite{richtler:99} \\
    1365 & 31.35 & 0.07 & Ceph & \cite{madore:98} \\
    \noalign{\medskip}
    \parbox{2.5cm}{1339, 1344, 1379,\\ 1399, 1404} & 31.37 & 0.12 
      & SBF & \cite{jensen:98} \\
    \noalign{\medskip}
    18 galaxies & 31.43 & 0.12 & T-F & 
      \cite{bureau:96}, \cite{rc3:91}, \cite{federspiel:98} \\
    \noalign{\smallskip}
    \hline
    \noalign{\smallskip}
    mean value & 31.35 & 0.04 & & \\
    \noalign{\smallskip}
    \hline
    \noalign{\medskip}
  \end{tabular}\\
  \setlength\tabcolsep{1ex}
  \begin{tabular}{rl}
    $^\mathrm{a}$~PNLF:&planetary nebulae luminosity function\\
    GCLF:&globular cluster luminosity function \\
    Ceph:&Cepheid distance; HST Key Project \\
    SBF:&surface brightness fluctuations \\
    T-F:&Tully-Fisher relation
  \end{tabular}
  \label{Fornax}
\end{table}

\section{The absolute magnitudes of type Ia SNe in Fornax 
  and the Hubble constant}

\index{supernova!absolute~magnitude} Table~\ref{tab:SNfornax} lists
the parameters of the three Fornax Ia events.  The photometry is taken
from \cite{hamuy:96}. For shortness, only the brightness $B$ is given.
SN\,1992A \index{SN\,1992A} in NGC\,1380 is one of the best ever
observed SNe. After correction for decline rate and colour and with
the new distance modulus of the Fornax cluster the absolute magnitudes
$M_\mathrm{corr}^B$ have been calculated.  Eq. (\ref{eq:h0}) and the
corresponding data in the $B$, $V$, and $I$ bands for the listed SNe
yield a Hubble constant \index{Hubble~constant} of
$H_0=72\pm6$\,km\,s$^{-1}$\,Mpc$^{-1}$.

\begin{table}
  \caption{Type Ia SNe in the Fornax cluster of galaxies}
  \label{tab:SNfornax}
  \begin{flushleft}
    \setlength\tabcolsep{4pt}
    \begin{tabular}{lllccccc}
      \noalign{\smallskip}
      \hline
      \noalign{\smallskip}
      SN & Host Galaxy &  $B_\mathrm{max}$ & $\sigma$ & $\Delta m_{15}$ 
        & $B_\mathrm{max}-V_\mathrm{max}$ & $B_\mathrm{corr}$ 
        & $M_\mathrm{corr}^B$ \\
      \noalign{\smallskip}
      \hline
      \noalign{\smallskip}
      SN\,1992A & NGC\,1380 &12.57 & 0.03 & 1.47 & 0.02 & 12.36 & $-18.99$\\
      SN\,1980N & NGC\,1316 &12.49 & 0.03 & 1.28 & 0.05 & 12.33 & $-19.02$\\
      SN\,1981D & NGC\,1316 &12.66 & 0.06 & 1.20 & 0.24 & 12.25 & $-19.10$\\
      \noalign{\smallskip}
      \hline
      \noalign{\smallskip}
    \end{tabular}
      Parameters of SN\,1981D from own light-curve fit
  \end{flushleft}
\end{table}

%INDEX%%%%%%%%%%%%%%%%%%%%%%%%%%%%%%%%%%%%%%%%%%%%%%%%%%%%%%%%%%%%%%%
\clearpage
\addcontentsline{toc}{section}{Index}
\flushbottom
\printindex
%%%%%%%%%%%%%%%%%%%%%%%%%%%%%%%%%%%%%%%%%%%%%%%%%%%%%%%%%%%%%%%%%%%%%

\end{document}